\documentclass[usenatbib,usegraphicx]{mn2e}
\title[Improvements for group delay fringe tracking]{Improvements for
group delay fringe tracking}
\usepackage{xspace}
\newcommand{\opd}{optical path difference (OPD)\xspace\renewcommand{\opd}{OPD\xspace}}
\author[A.G. Basden et al.]{A.G.~Basden$^{1,2}$\thanks{E-mail:
abasden@mrao.cam.ac.uk} and D.F.~Buscher$^1$\\
$^1$Astrophysics Group, Cavendish Laboratory, Madingley Road,
Cambridge CB3 0HE\\
$^2$Astronomical Instrumentation Group, Department of Physics, Science
Laboratories,\\ University of Durham, South Road, Durham, DH1 3LE\\}

\begin{document}
\date{Released 2004 Xxxxx XX}

\pagerange{\pageref{firstpage}--\pageref{lastpage}} \pubyear{2003}
\label{firstpage}
\maketitle

\begin{abstract}
Group delay fringe tracking using spectrally-dispersed fringes is
suitable for stabilising the optical path difference in ground-based
astronomical optical interferometers in low light situations.  We
discuss the performance of group delay tracking algorithms 
when the effects of atmospheric
dispersion, high-frequency 
atmospheric temporal phase variations, non-ideal path
modulation, non-ideal spectral sampling, and the detection artifacts
introduced by electron-multiplying CCDs (EMCCDs) are taken into
account, and we present ways in which the
tracking capability can be optimised in the presence of these effects. 
\end{abstract}

\section{Introduction}
A stellar optical interferometer combines the light collected from a
star by different telescopes to produce interference fringes. In order
to see high-contrast fringes, the total optical path travelled by the
light from the star to the plane of interference must be the same to
within a fraction of the coherence length of the light, regardless
which telescope the light beam has passed through.  In ground-based
interferometers the light arriving through different telescopes is
subject to different and randomly-varying optical path perturbations
caused by atmospheric and instrumental effects. The resulting \opd is
potentially many times larger than the coherence length, and so most
interferometers incorporate an active method of compensating for this
\opd error in real time.

The usual compensation method adopted is to sense the \opd error using
some property of an interference pattern formed using light from the
target star or a nearby reference star and to actuate a variable path
delay in one arm of the interferometer to compensate for this
error. This process is known as ``fringe tracking''.  Three major
techniques for fringe tracking are used in interferometers worldwide:
phase tracking making use of phase unwrapping methods
\citep{phasetracking}, envelope tracking by monitoring the fringe
coherence envelope \citep{envelopetracking} and group delay fringe
tracking (hereafter GDFT) by the analysis of spectrally dispersed
starlight fringes \citep{groupdelaytracking}.  GDFT has the advantage
over the other methods mentioned in that it is able to successfully
track fringes at lower light levels \citep{buscher,lawson}. The
performance of the fringe tracker at low light levels typically sets
the magnitude limit of the interferometer \citep{buscher} and so is a
critical performance parameter of an interferometer system
design. This paper concentrates only on the GDFT method.

Some limitations of GDFT have been studied by \citet{lawson} and
references therein.  However these studies have not considered a
number of important problems common to real interferometers.  In this
paper, we quantify the effects of some of the instrumental factors
that can affect the tracking capability of a system. These factors include
atmospheric longitudinal dispersion, temporally-varying atmospheric
phase perturbations, non-ideal spectral sampling, non-ideal optical
path modulation, and the detector characteristics typical of
electron-multiplying CCDs (EMCCDs).  We also determine how the values
of a number of parameters of the GDFT data-processing algorithm,
including the window function selection, delay sampling interval and
spectral channel binning factor can improve or degrade the tracking
capability of a GDFT system.

The Cambridge Optical Aperture Synthesis Telescope (COAST) is being
upgraded to allow fast spectro-interferometric fringe detection and
group delay fringe tracking. In this paper, we use ranges of values
for various system parameters which are relevant to the design of the
GDFT system on the COAST interferometer, but the results we present
are likely to be relevant to other interferometers worldwide.

In \S2 we introduce our model for the interferometer, including the
fringe detection model and the non-ideal properties of real
group-delay tracking systems which we shall be considering.  In \S3 we
describe the basic group delay tracking algorithm and the
modifications to the GDFT algorithm which can be used to increase the
performance of the algorithm under these non-ideal conditions.  In \S4
we explain the numerical simulations used to test these modifications
and in \S5 we provide a discussion of the results.  We present our
conclusions in \S6.

\section{Instrument model}
We introduce here a model of the fringe detection process which
attempts to be as general as possible, so that it can be applied to
different interferometers. At the same time we will restrict ourselves
to values of model parameters which are appropriate to the
COAST interferometer so as to give a concrete example of the model and
also to simplify the discussion.

\subsection{Fringe measurement}
Interference patterns in optical interferometers are generally
observed in the form of quasi-sinusoidal intensity variations. In
different interferometers these variations may be detected spatially,
for example in image-plane beam combination schemes, or temporally as
exemplified by path-modulated pupil-plane schemes. We treat here only
a temporally-sampled detection scheme, but most of our results will be
applicable to a spatially-sampled scheme, since the mathematical
treatment of these two sampling methods is similar. 

We explicitly treat only single-baseline fringe-measurement
schemes. Generalisation to multi-baseline schemes is straightforward
in most cases: typically separate beam combiners are used to measure fringes on
different baselines or an orthogonal encoding scheme is used to
multiplex fringes from different baselines onto a single detector (for
example by making fringes from different baselines appear at different
temporal frequencies), and so analysis of the fringes on different
baselines is a separable problem. We do not consider here algorithms which make
use of the correlation of the information available on different baselines.

In typical temporally-sampled schemes, the optical path delay is
modulated as a function of time (using for example a rapidly-moving
mirror) with an approximately sawtooth pattern , i.e.\ each modulation
cycle consists of two segments where the \opd between the two
interfering beams is made to change approximately linearly as a
function of time, with the rate of change of \opd changing sign
between segments. We call a single segment of this modulation pattern
a ``sweep''.

\begin{figure}
(a)\includegraphics[angle=270,width=8cm]{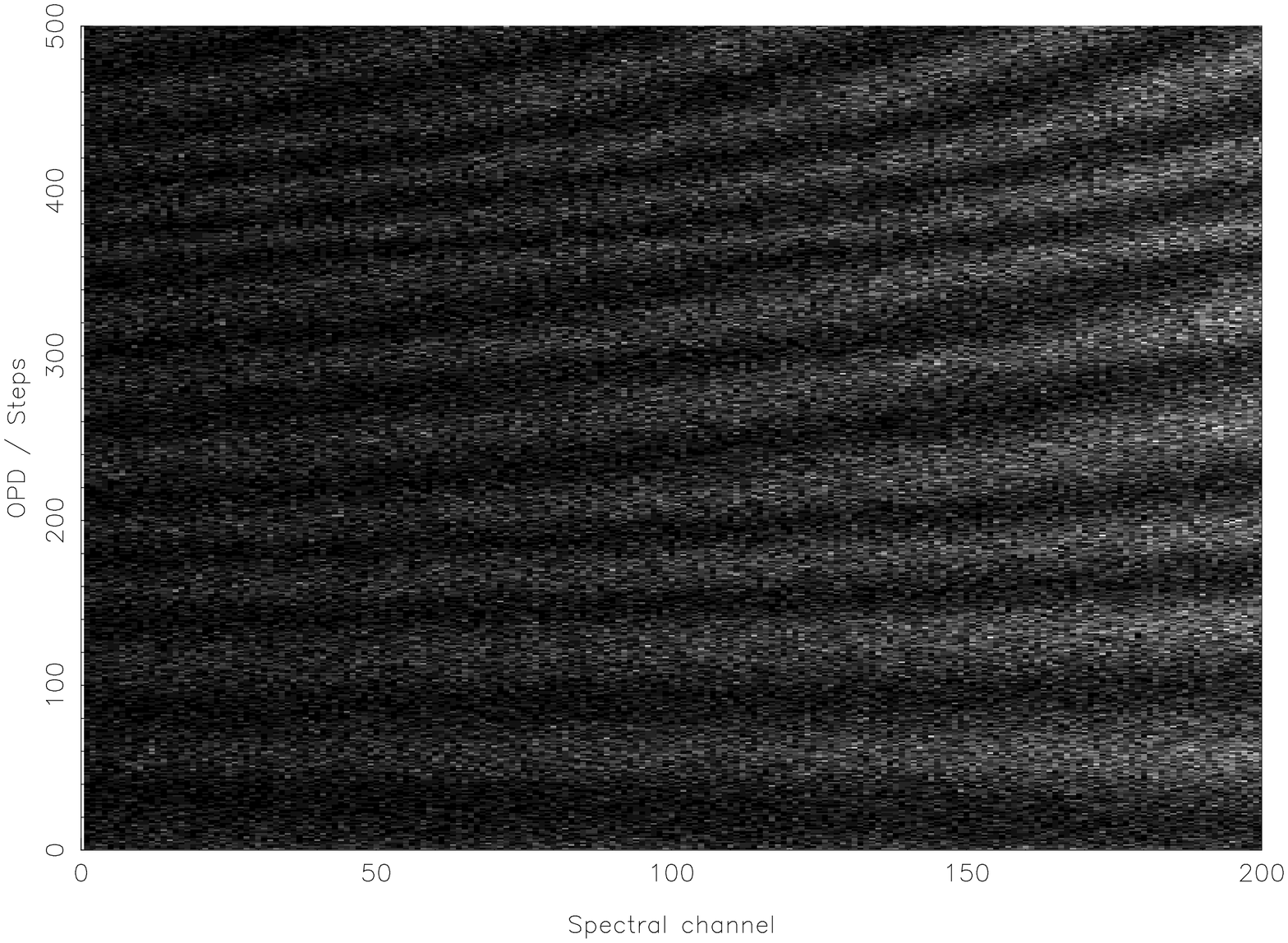}
(b)\includegraphics[angle=270,width=8cm]{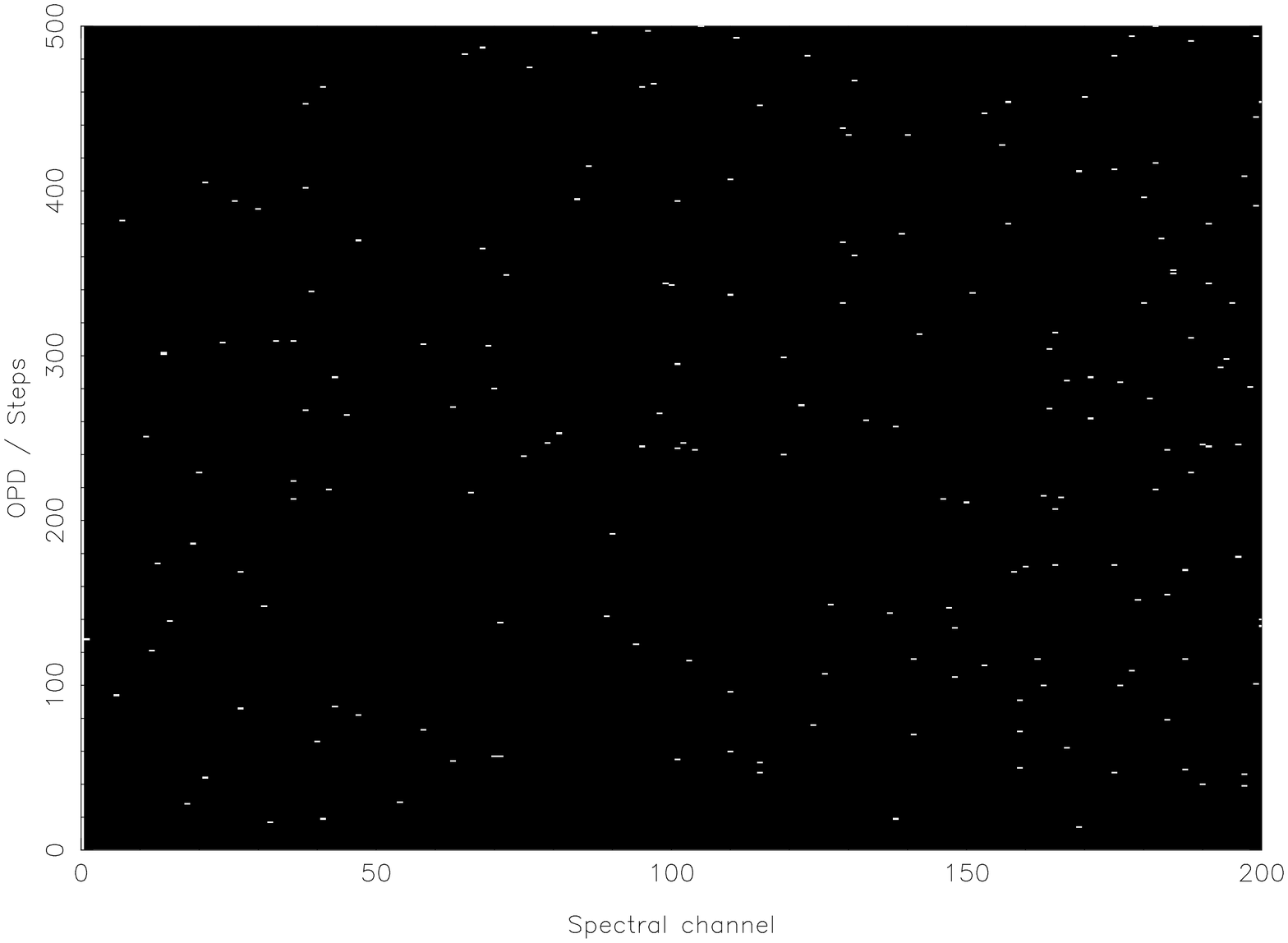}
\caption{(a) Example of an interferometric fringe pattern showing
fringes with a visibility amplitude of 0.5 and a mean light level of
20000 photons per atmospheric coherence time (2 photons per pixel),
dispersed using an SF18 glass prism, and with a linear sweep.  A
vertical cross section of the fringe pattern will yield an
quasi-sinusoidal intensity variation with the \opd (time).  The total
distance scanned in the \opd (y) axis is 7.6~$\mu$m, and the effect of
atmospheric fluctuations can be seen to cause a change in fringe
period with time.  (b) A fringe identical to that in (a), but with a
mean light level of only 20 photons per atmospheric coherence time
(0.002 photons per pixel).  The mean tracking capability for this
signal level is approximately 0.75, even though no fringe is visible
to the human eye.}
\label{fig:fringepatternpics}
\end{figure}

Sampling the flux emerging from the beam combiner with a single
detector yields a one-dimensional sequence of intensity values in which
the fringe pattern appears as a quasi-sinusoidal change of the
intensity with time.  If the flux is split into a number of spectral
channels, for example by dispersing the light with a prism or grism
onto a one-dimensional array of detector pixels, then the set of
intensity values measured can be represented as a discrete
two-dimensional function, one coordinate being the time at which the
intensity was sampled and the other being the detector pixel
number. We shall call a ``data frame'' the two-dimensional dataset
obtained during the course of a single sweep. Simulated data frames at
different light levels are shown in figure~\ref{fig:fringepatternpics}.

The intensity of an 
element of such a data frame (adapted from \citet{scott}) can
be written as:
\begin{eqnarray}
\label{fringeintensity}
I_{jk} & = & I_m(t_k, \sigma_j) + I_n(t_k, \sigma_j)
\nonumber \\
&& + 2V_{mn}(t_k, \sigma_j)\sqrt{I_m(t_k, \sigma_j)I_n(t_k, \sigma_j)} 
\nonumber\\ 
&&\;\;\times \cos \left[ 
2\pi\sigma_j \left(
l_\mathrm{mod}(t_k)+l_\mathrm{atm}(t_k) + l_\mathrm{static}
\right) 
\right. 
\nonumber \\
&& \;\;\;\;\left.
+ \phi_{mn}(\sigma_j) + \theta_{mn}(\sigma_j)
\right],
\end{eqnarray}
where 
$I_{jk}$ is the intensity measured on pixel $j$ of the detector array 
at time $t_k$,
the mean wavenumber of the light falling onto
pixel $j$ is given by $\sigma_j$ (i.e. $1/\lambda_j$),
 $I_m$ and $I_n$ 
are the intensities of
the light arriving from telescopes $m$ and $n$ respectively, 
and $V_{mn}$ represents the fringe visibility modulus.
This visibility modulus takes into account the modulus of the source coherence
function at wavenumber $\sigma_j$ on the baseline between telescopes
$n$ and $m$, the effects of ``smearing'' of the fringes caused by
fringe motion during the integration period corresponding to sample
$t_k$, the effects of fringe smearing due to fringe phase shifts
across the finite spectral bandpass falling on detector pixel $j$, and
the effects of differential spatial aberrations (both static and
randomly-varying) across the interfering wavefronts from telescopes
$m$ and $n$.

The fringe phase is split into \opd terms, which give rise to phase
shifts which scale linearly with wavenumber, and other phase terms.
The total \opd is made up of three terms: $l_\mathrm{mod}(t_k)$
represents the \opd variation deliberately introduced by the path
modulation system, $l_\mathrm{atm}(t_k)$ represents the random \opd
fluctuations due to the atmosphere and $l_\mathrm{static}$ is a
``static'' \opd offset which remains fixed for the duration of the
sweep. The offset $l_\mathrm{static}$ includes any instrumental delay
plus the mean atmospheric \opd during the sweep, and it is this
quasi-static \opd which is the quantity to be estimated by our
fringe-tracking algorithm.

The phase of the source coherence function at wavenumber $\sigma_j$
when measured on the baseline between telescopes $m$ and $n$ is
written as $\phi_{mn}(\sigma_j)$, and $\theta(\sigma_j)$ is a
wavelength-dependent phase shift due to either instrumental effects or
the wavelength dependent refractive index of the atmosphere. It is to
be noted that $l_\mathrm{static}$ can be taken to include any component of the 
quasi-static differential 
phase shifts which changes linearly with wavenumber,
so that
$\theta(\sigma_j)$ typically represents only the non-linear component of the
combined instrumental and atmospheric dispersion.

For the rest of this paper we ignore the effects of atmospheric
scintillation, so that $I_m$ and $I_n$ are constant in
time, and we further assume that $I_m(\sigma_j)=I_n(\sigma_j)=I_0(\sigma_j)$.
We also assume (with little loss of generality) 
that the fringe intensity is sampled at equally-spaced
intervals of time, i.e.\ $t_k$ is a linear function of $k$.

\subsection{Detector model}
\label{l3ccdsect}
The requirements on a detector to be used for group delay fringe
tracking are stringent.  Since we wish to record many spectral
channels simultaneously, an array of detector elements is required.  One
solution is to use an array of avalanche photo-diodes (APDs), as used
for example at the Navy Prototype Optical Interferometer (NPOI) \citep{npoi}.  However, this solution limits the number of spectral
channels that can be used, since APDs are relatively expensive and fragile,
and require fibre feeds if many are to be used, reducing the optical
throughput.

Another solution is to use a charge coupled device (CCD). These devices
have traditionally not been used extensively in fringe detection due to the
high readout noise at the fast (MHz) pixel rates required.
However, electron multiplying
CCDs (EMCCDs) \citep{mackay} with sub-electron readout noise
have recently been made available, and these are
an attractive detector option for GDFT systems
since they retain all the positive
properties of conventional CCDs (such as low dark current when cooled,
stability, linearity).  Using these devices, we can achieve a
combination of high
spectral resolution, high temporal resolution, high quantum efficiency and
photon-counting performance, allowing fringe
detection at much lower signal levels than has previously been
possible.

We consider here EMCCDs as detectors for group delay fringe
trackers. With sufficiently high on-chip gain and at low photon rates,
EMCCDs can be considered as ideal photon-counting devices. More
detailed modelling needs to take into account the additional noise in
the EMCCD output signal due to the stochastic on-chip gain process,
which can reduce the output signal-to-noise ratio (SNR) by up to a
factor of $\sqrt{2}$ compared with the photon-noise-limited SNR. We
use here the noise model described in \citet{basden} and
\citet{basden2}, and the algorithms described therein, which can be
used to improve the SNR of the fringe signal.

\subsection{Spectrometer model}
\begin{figure}
\includegraphics[width=8cm]{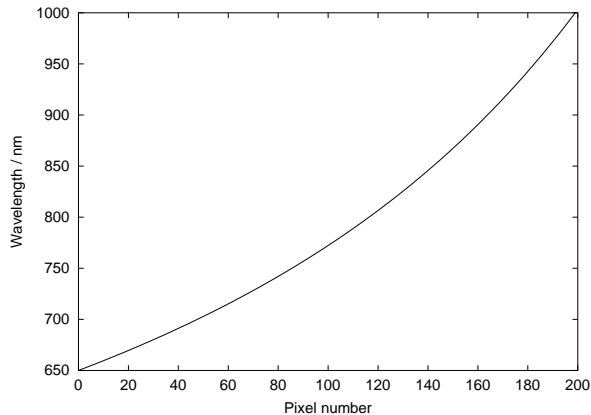}
\caption{The dispersion relationship used to simulate dispersion from
an SF18 equilateral prism, with a wavelength range of 350~nm.}
\label{fig:prismdisp}
\end{figure}

Previous analyses of GDFT algorithms have assumed that fringe pattern
is spectrally dispersed
such that $\sigma_j$ is a linear function of $j$ so that Discrete
Fourier Transform (DFT) algorithms can be used.
This ideal can be approached by using a grism and a uniformly-spaced
linear array of detector pixels
\citep{traub}. The GDFT system at COAST uses an SF18 glass prism for
the dispersion and a linear array of EMCCD pixels, 
and this gives a roughly quadratic relationship
between the pixel number and the wavenumber, as shown in
figure~\ref{fig:prismdisp}. A second alternative would be to use a
diffraction grating, which gives a spectral dispersion which is linear
in wavelength, i.e. inversely proportional to wavenumber. 

The wavelength range shown in figure~\ref{fig:prismdisp} is similar
to that sampled by the COAST GDFT system, i.e.\ 650-1000nm, sampled
with 200 detector pixels.

\subsection{Path modulation}
On the COAST interferometer, the fringe
intensity is typically sampled at a rate of 5 kHz, 
and a single \opd modulation sweep will last for 100~msec, i.e.\ 500
samples. The modulation $l_\mathrm{mod}(t)$ 
is an approximate sawtooth in shape with 
a peak-to-peak path-length change of 59~$\mu$m.  The interference signal for
a wavelength of
800nm therefore appears as a sinusoidal modulation
with a fringe period of 6.56 samples. 

The finite acceleration of the path modulator on COAST means that the
sawtooth has ``rounded edges'' which last for approximately 5-10\% of
the sweep. Thus the fringe period will change slightly at the edges of
the sweep.

\subsection{Atmospheric phase perturbations}
The random atmospheric fluctuations $l_\mathrm{atm}(t)$ are assumed to
arise from atmospheric turbulence above the telescopes following a
Kolmogorov-Tatarski model and the Taylor ``frozen-turbulence''
hypothesis \citep{tatarski}. This model yields 
the temporal structure function of phase ($\phi$) variations with time
at a single point $r$ on the wavefront as:
\begin{equation}
D_\phi (t) = \left< | \phi(r, t' +t)-\phi(r, t')|^2\right> =
\left(t/t_0\right)^{5/3}
\end{equation}
The atmospheric fluctuations are to a good approximation pure delay
fluctuations, so the
coherence time, $t_0$, scales with wavelength as $\lambda^{6/5}$.
On short time-scales we can
assume the fluctuations at any two telescopes are uncorrelated, so
that the structure function of the delay \emph{difference} is just
twice the structure function of the delay variation at a single
telescope, i.e.
\begin{equation}
\left< | l_\mathrm{atm}( t' +t)-l_\mathrm{atm}(t')|^2\right> = \lambda_0^2
\left(t/t_0\right)^{5/3}/(2\pi^2)
\label{kolmogorov}
\end{equation}
where $\lambda_0$ is the wavelength at which $t_0$ is defined. Typical
values of $t_0$ at COAST  are around 10~msec at a wavelength of
0.8$\mu$m, i.e. the sweep is approximately 10$t_0$ in length.

\subsection{Atmospheric longitudinal dispersion}
\begin{figure}
\includegraphics[width=8cm]{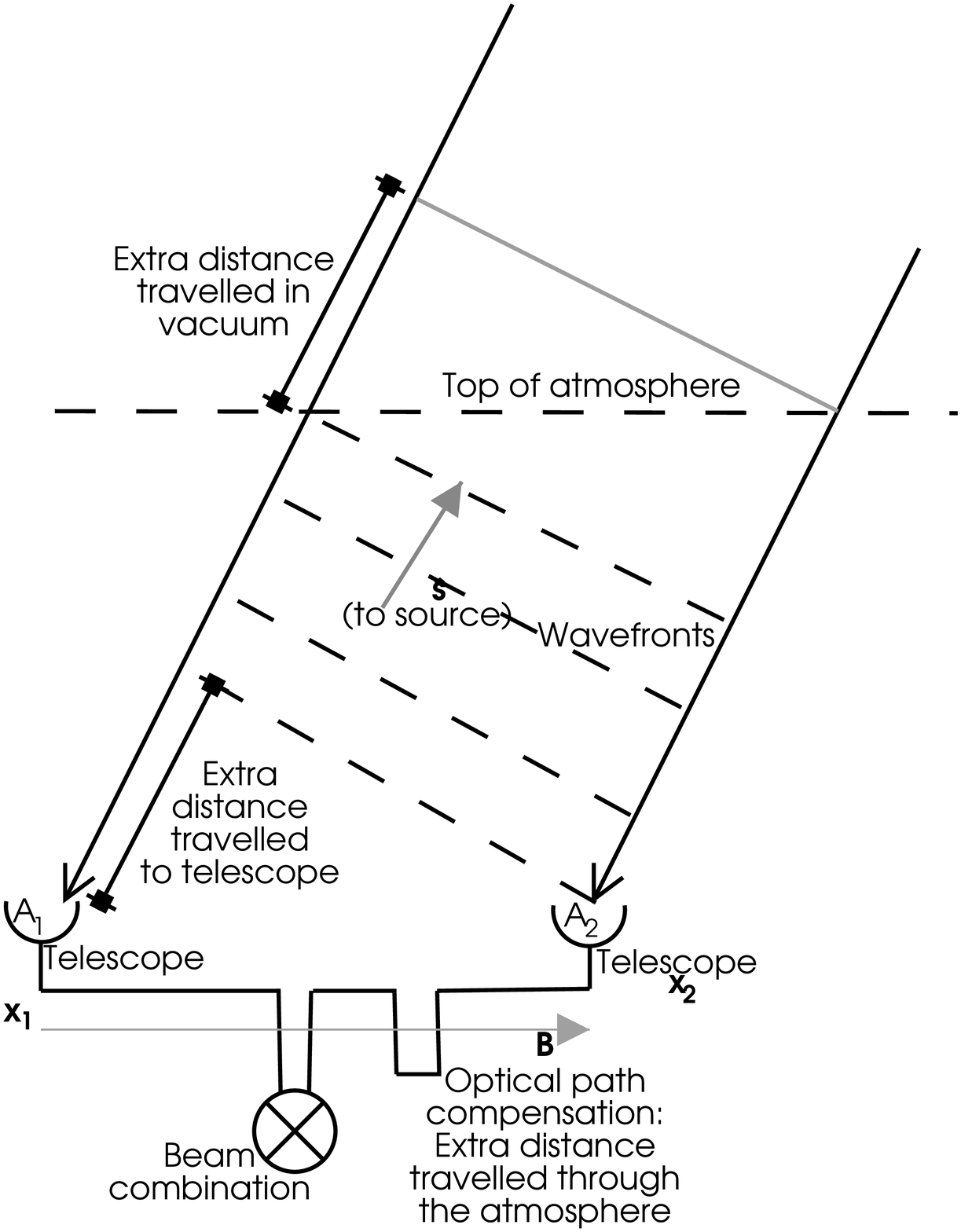}
\caption{A demonstration of how array orientation causes light to
travel different distances through the atmosphere even when the \opd is
zero, if optical path compensation is not carried out in a vacuum,
between two telescopes at positions $x_1$ and $x_2$, separated by a
baseline B.}
\label{opddiagram}
\end{figure}

The refractive index of air, $\eta(\sigma)$, is dependent on the
wavelength of light travelling through it.  If \opd compensation is not
carried out in a vacuum, 
it is quite possible for the total path travelled in air by beams from
two different telescopes to differ by many tens or hundreds of 
metres as shown in
figure~\ref{opddiagram}.

The wavelength dependence of the atmospheric refractive index
introduces a phase difference between light beams which have travelled
through different amounts of air.
This longitudinal
dispersion (relative to a reference wavenumber $\sigma_\mathrm{ref}$)
can be written by performing a Taylor series expansion on the
expression given by \citet{cox} as
\begin{eqnarray}
\theta_{mn}(\sigma_j,\sigma_{ref}) &=& 2\pi \sigma_j D_{mn}
\left(1-\frac{\eta\left(\sigma_{ref}\right)}{\eta\left(\sigma_j\right)}\right)\\
&\approx & 2\pi \sigma_j D_\mathrm{air} \left[
\left(\sigma_j^2 -\sigma_{ref}^2\right) \right. \nonumber \\
&&\times \left( 1.5358\frac{p T_s}{p_s T} +0.346\frac{p_w}{p_s} \right) \nonumber \\ 
&& +\left(\sigma_j^4 - \sigma_{ref}^4 \right)\left(1.318\times 10^4\frac{p
T_s}{p_s T} \right) \nonumber \\ 
&& +\left. \left(\sigma_j^6 - \sigma_{ref}^6 \right)\left(1.55\times
10^8\frac{p T_s}{p_s T} \right)\right] \label{eq:longitudinal} \nonumber\\
\end{eqnarray}
where $D_\mathrm{air}$ is the net difference in air path travelled by
the two beams. For an interferometer with an air delay line (such as
COAST) this is typically equal to $D_{nm}$, the projection of the
baseline vector of the telescope pair $n, m$ onto the source
direction.  The atmospheric pressure is $p$, with $p_w$ the water
vapour partial pressure and $p_s=0.101325$~MPa.  The atmospheric
temperature is $T$ with $T_s=288.15$~K.  We take the value of
$\sigma_\mathrm{ref}$ to be 1/650~nm$^{-1}$, though this does not
constrain the calculations in any way.

\begin{figure}
\includegraphics[width=\linewidth]{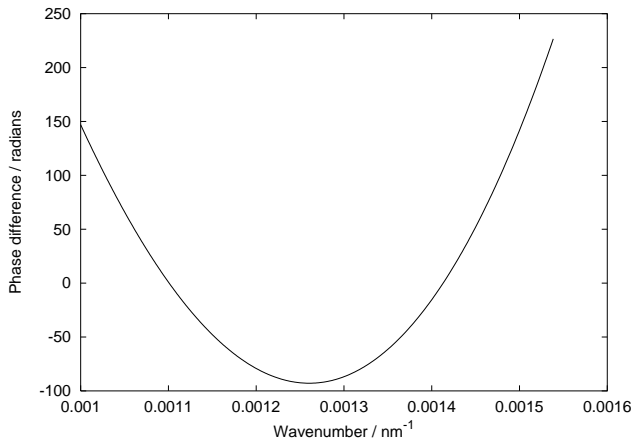}
\caption{The effect of the longitudinal dispersion of the atmosphere
  for a system with air delay lines.  The graph shows the residual
  phase shift as a function of wavelength over the wavelength range
  650-1000nm for 100m of net air path difference.  The
  best-fit group delay (i.e. linear trend of phase with wavenumber)
  has been removed.}
\label{fig:longitudinalexample} 
\end{figure}

It is possible for the phase difference introduced by longitudinal
dispersion to be of order several hundred radians, as shown in
figure~\ref{fig:longitudinalexample}.

\section{OPD estimation algorithm}
\label{sec:opdestimation}

\begin{figure}
(a)\includegraphics[angle=270,width=\linewidth]{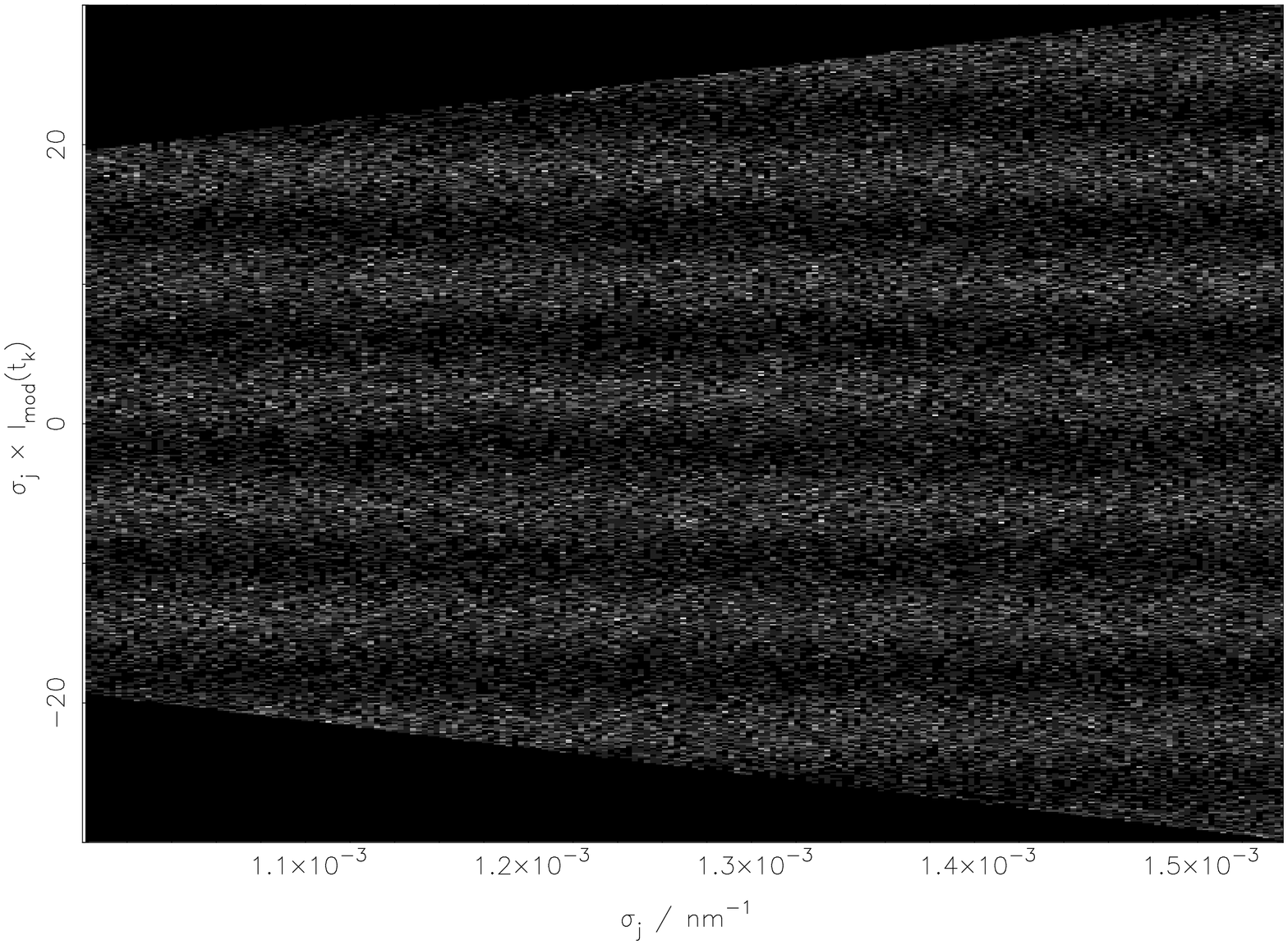}
(b)\includegraphics[angle=270,width=\linewidth]{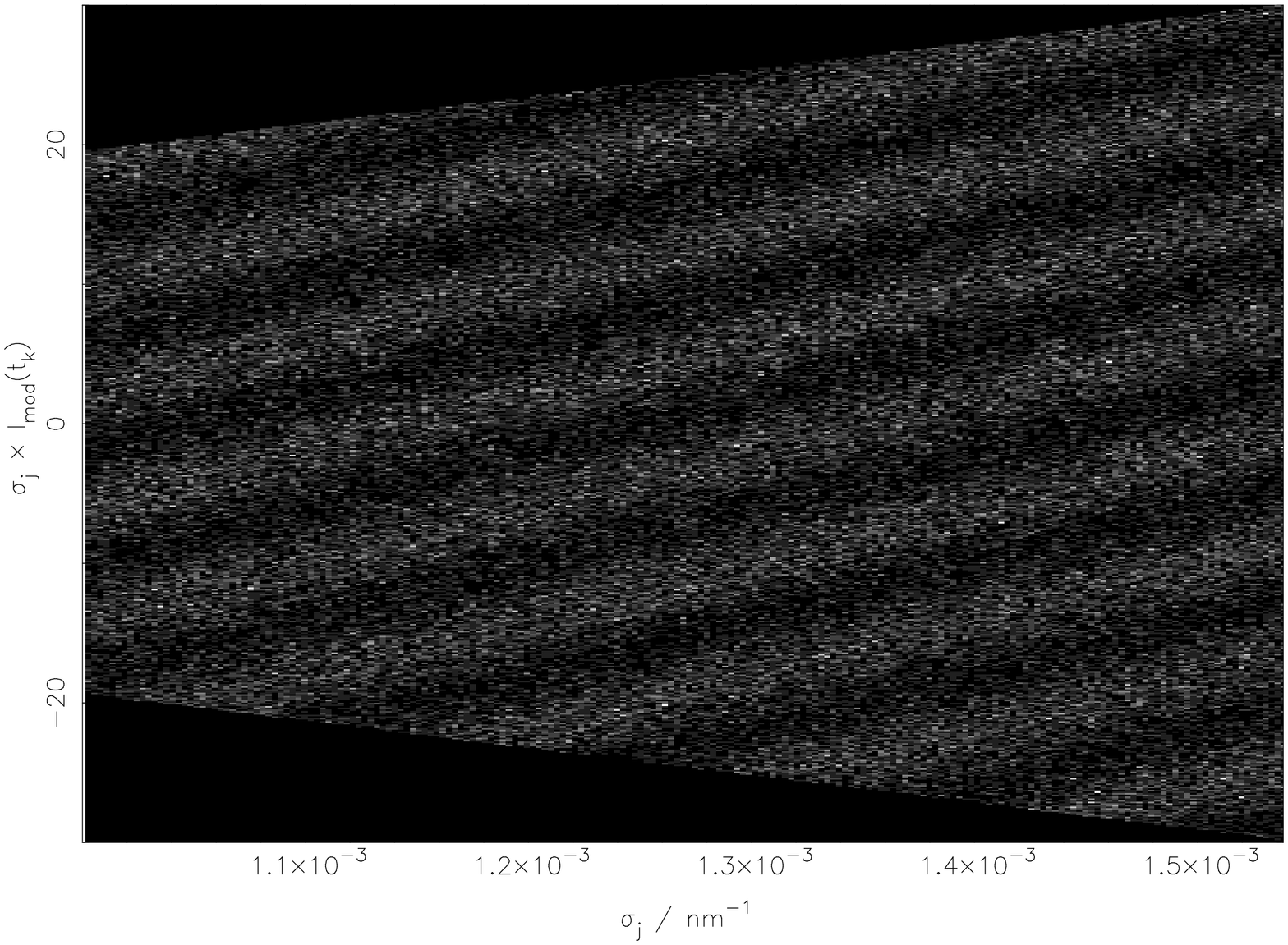}
\caption{Fringe data with a visibility amplitude of 0.5 and mean light
level of one photon per pixel plotted in a coordinate system with the
$y$ coordinate scaled by the wavenumber $\sigma_j$.  The black areas
outside the trapezoidal region containing fringes represent regions
for which there is no data. (a) Showing the fringe pattern when the
\opd is zero.  (b) Showing the fringe pattern when the \opd is
non-zero.}
\label{fig:rescaledfringe}
\end{figure}

In an idealised interferometer where there is no atmospheric or
instrumental dispersion, where the sweep time is much faster than the
time-scale of atmospheric phase fluctuations, and where the source
phase changes only slowly with wavelength, we can set
$\phi_{mn}(\sigma_j),\theta_{mn}(\sigma_j) =\mathrm{constant}$ and
$l_\mathrm{atm}(t)=0$ in Eq.~\ref{fringeintensity}. If we were then
to plot the data frame as a function of $\sigma_j$ and $\sigma_j
l_\mathrm{mod}(t_k)$, as shown in Fig.~\ref{fig:rescaledfringe}, it can be
seen that the measured intensity forms a sinusoidal pattern in the
two-dimensional plane such that the lines of maximum and minimum
intensity are sloped by an amount proportional to
$l_\mathrm{static}$. That is to say, when $l_\mathrm{static}$ is zero,
the phase of the fringes is constant with wavelength and so the lines
of fringe maxima lie parallel to the $\sigma_j$ axis, whereas when
$l_\mathrm{static}$ is non-zero, the phase changes linearly with
wavenumber, leading to a slope in the locus of fringe maxima.

The above argument indicates that we can determine
$l_\mathrm{static}$, the \opd offset, by measuring the rate of change
of fringe phase with wavenumber $\sigma$.  This process can be
implemented using a two-dimensional Fourier transform in the case of
an idealised interferometer, and this approach has been used by other
authors (\citet{lawson} and references therein). Here we describe a
more generalised two-step process which is equally applicable to an
ideal or a non-ideal interferometer.

In the first step, we estimate the fringe amplitude and phase (i.e.\
the complex fringe amplitude) for each spectral channel.  For data
which is sampled at equal intervals of path delay and over an integral
number of fringe cycles, this process is identical to performing a
one-dimensional DFT on the time series of
intensities in each wavelength channel in order to extract a single Fourier
component corresponding to the fringe frequency at that wavelength.

In the second step, we choose a set of different trial \opd offsets
$\{l_p\}$.  For each trial \opd $l_p$, we shift the phase of the
complex amplitude in each spectral channel by an amount $-2\pi
l_p\sigma_j$ (where we note that the phase shift is a linear function
of wavenumber), sum the resulting complex amplitudes over all spectral
channels and take the squared modulus of the result. This squared
modulus will typically be maximised when
$l_p=l_\mathrm{static}$. Thus the value of $l_p$ which gives a peak in
peak in the squared
modulus gives an estimate of the \opd error. 
For data which is sampled at equal intervals of wavenumber, this
second step is identical to looking for a peak in the power spectrum
of the 1-dimensional complex amplitude data.

We now describe the two steps of the \opd estimation algorithm in more detail.

\subsection{Determining fringe amplitude and phase}
In a perfect interferometer where $l_\mathrm{mod}(t_k)$ is a linear
function of $k$ during a sweep, 
the complex fringe amplitude can be estimated using
a simple DFT. We
use here a modification of the DFT algorithm which can be used in
cases of moderate non-linearity of the modulation waveform. An
algorithm applicable over a wider range of modulation non-linearity
is discussed in \citet{thorsteinsson}.

We estimate the complex fringe amplitude $F_1(\sigma_j)$ in
spectral channel $j$ using a weighted sum over the time axis of the data frame:
\begin{equation}
F_1(\sigma_j)  =  \sum_{k=1}^N W_1(t_k)\exp[-2\pi i\sigma_j
  l_\mathrm{mod}(t_k)]\left.{\rm d t \over \rm d l_\mathrm{mod}(t)}\right|_{t_k}I_{jk}
\label{eq:first-dft}
\end{equation}
where $N$ is the number of intensity samples in a sweep, and
$W_1(t_k)$ is a real-valued weighting function.  Multiplying the data
by $\exp[2\pi \sigma_j l_\mathrm{mod}(t_k)]\left.{\rm d t \over \rm d
l_\mathrm{mod}(t)}\right|_{t_k}$ and summing the result over $k$ is
approximately equivalent to transforming the measured intensities to a
coordinate system whose independent coordinate is $l_\mathrm{mod}$
rather than $t$ and then performing a Fourier transform.  The
coordinate transformation serves to remove many of the effects of the
non-linearity of $l_\mathrm{mod}$ as a function of $t$.

The weighting function $W_1(t_k)$ is a ``window function'' which
typically tapers towards the edge of the sweep. The window function is
introduced to reduce so-called ``spectral leakage'' effects, which
cause components of the fringe pattern at frequencies different from
the fringe frequency to contaminate the fringe signal. Typically
spectral leakage becomes important when the number of fringe periods
in a sweep is non-integral (see \citet{numrecipe} for further
details).  Since there is no single ideal windowing function we
investigate some given by \citet{numrecipe} (the Welch, Bartlett,
Hann, Hamming and Blackman windows), as well as the default ``top-hat''
window.

Note that equation~\ref{eq:first-dft} becomes identical to evaluating
the DFT of the fringe intensity $I_{jk}$ if we set $W_1(t_k)=1$ and
$l_\mathrm{mod}$ is a linear function of $t$.

\subsection{Estimating the fringe power as a function of
  delay offset}
\label{finaldataarray}
When determining the total fringe amplitude $F_2$ 
as a function of the trial group
delay $l_p$, we
again use a generalised form of the DFT to
account for a possibly non-linear variation of wavenumber with pixel
index:
\begin{equation}
F_2(l_p)  =  \sum_{j=1}^M W_2(\sigma_j)\exp(-i\theta_j)\exp(-2\pi i\sigma_j l_p)\left.{\rm d x \over \rm d \sigma}\right|_{\sigma_j}F_1(\sigma_j)
\end{equation}
where $W_2$ is a window function with a similar form to that used for
$W_1$. The $x$ coordinate in the differential is a continuous extension
of the discrete spectral pixel coordinate $j$ --- in the typical case of a spectrometer
having a uniform linear array of square detector pixels, the $x$ coordinate is
a scaled version of 
the spatial coordinate running along the long axis of the array, with
$x=j$ at the centre of pixel $j$.
The factor $\exp(-i\theta_j)$ serves to compensate for any
wavelength-dependent phase variations
due to the dispersion of the atmosphere or
to instrumental effects. We typically assume that $\theta_j$ can be
computed with sufficient accuracy using calibration measurements
and/or atmospheric models, although we also investigate later the
effect of using inaccurate atmospheric models.

We choose to evaluate
$F_2$ over an evenly-sampled range of delays $l_p$ such that:
\begin{equation}
l_p = p s/\Delta\sigma, \quad p=\left\{-N_p+1, -N_p+2, \ldots N_p\right\}
\label{eq:delayscaling}
\end{equation}
where $\Delta\sigma$ is the wavenumber bandpass of the spectrometer
$\Delta\sigma=\sigma_1-\sigma_M$ (the difference between the minimum
and maximum wavenumbers), $s$ is a scaling factor of order unity, and
$2 N_p$ trial delays are sampled. The value of $s$ can be adjusted to
sample the group more or less finely, giving a trade off between
resolution and \opd range sampled for a given value of $N_p$.  The
value of $N_p$ can be adjusted to sample a larger or smaller range of
possible \opd offsets.

The resolution (smallest \opd identifiable) is given by
$s/\Delta\sigma$.  The range of delay space sampled is given by
$N_ps/\Delta\sigma$.  With unlimited computing power, $N_p$ could be
made as large as desired with $s$ kept small, though in practice there
is little point in sampling beyond the coherence length of the
individual spectral channels, since the signal here is much reduced.
If a constant range of delay space is to be sampled, the number of
sample points must therefore depend on the scale factor.

\begin{figure}
(a)\includegraphics[angle=270,width=\linewidth]{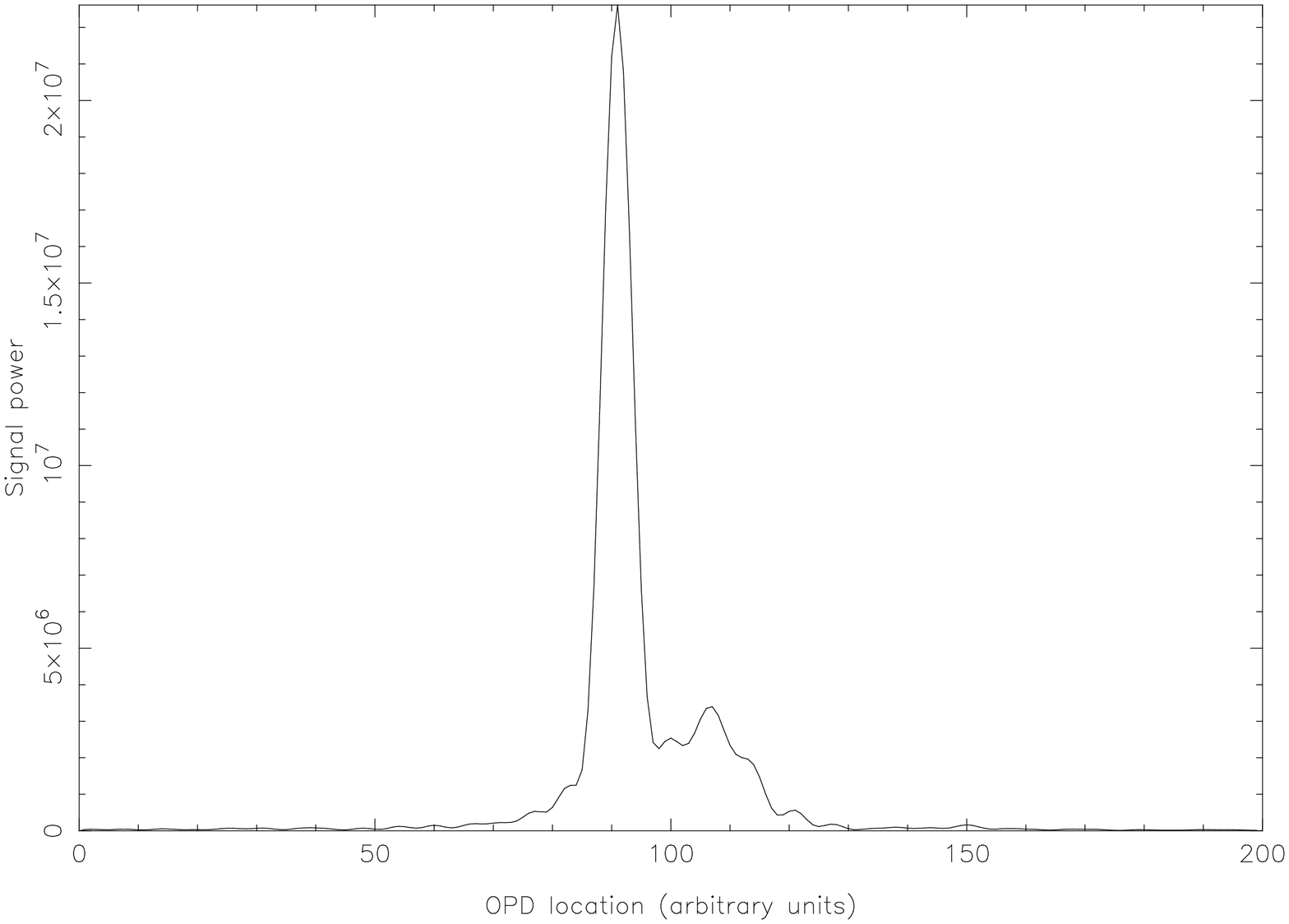}
(b)\includegraphics[angle=270,width=\linewidth]{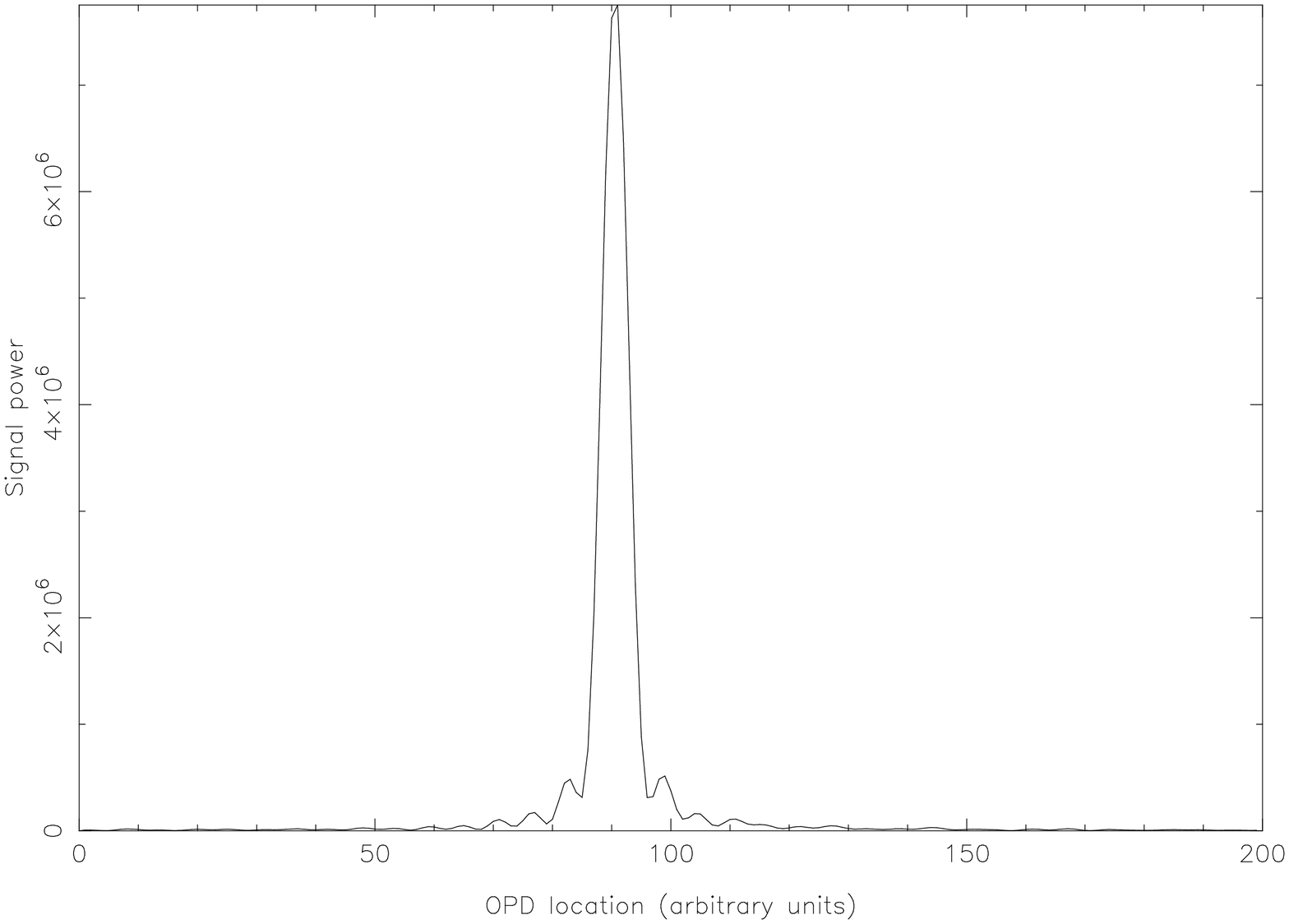}
(c)\includegraphics[angle=270,width=\linewidth]{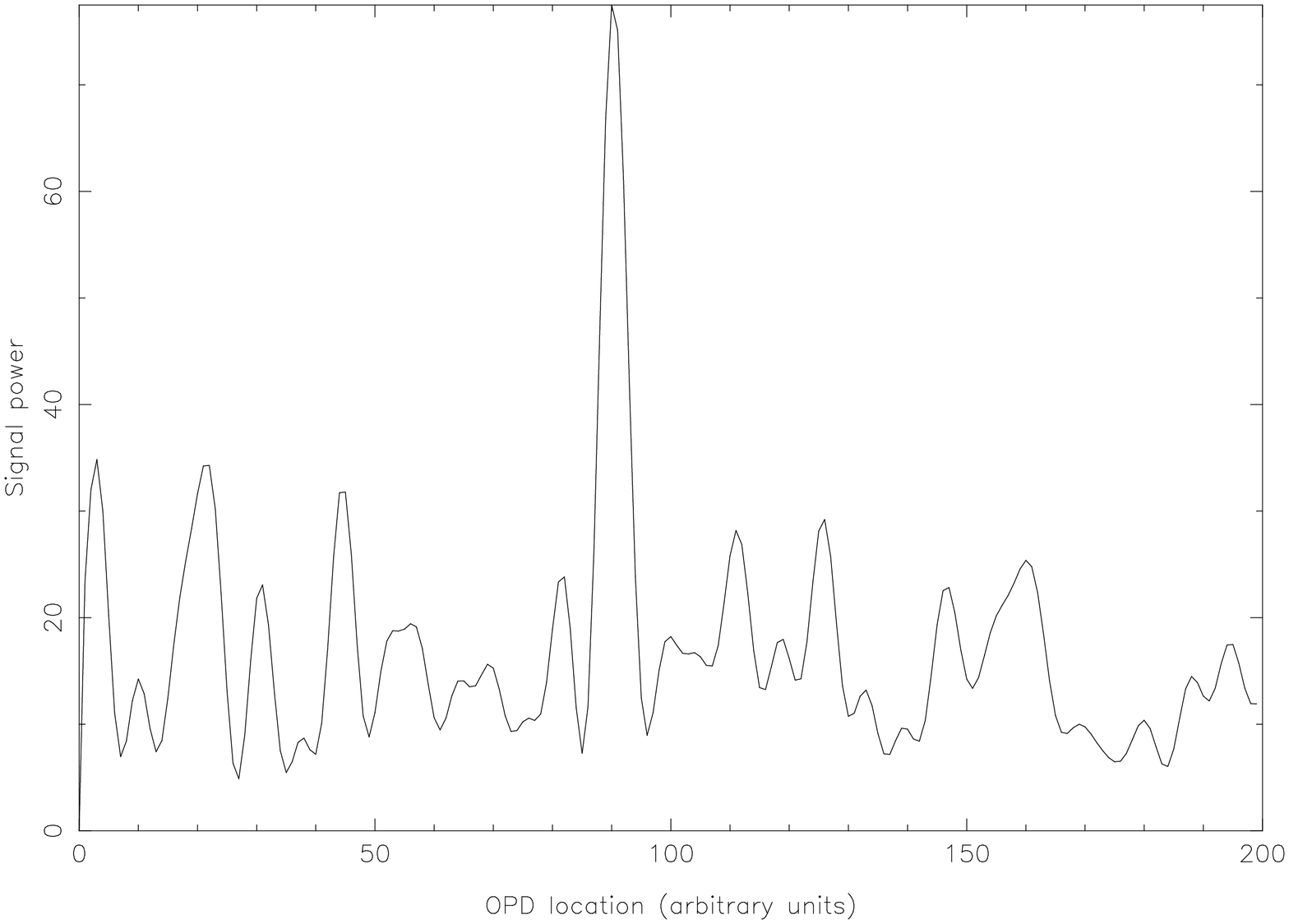}
\caption{Examples of a typical result from the OPD locating algorithm.
(a) shows the group delay signal obtained from the fringe pattern
Fig.~\ref{fig:fringepatternpics}(a) using top-hat windowing
functions, and spectral leakage is obvious.  (b) shows the group delay
signal obtained from a fringe pattern created using identical
parameters as Fig.~\ref{fig:fringepatternpics}(a) using a
Bartlett window during the first stage of the \opd estimation
algorithm.  This helps to demonstrate the effect that windowing
functions have, removing spectral leakage.  Plot (c) shows the signal
obtained from the fringe patterns
Fig.~\ref{fig:fringepatternpics}(b), with top-hat windowing
functions.  The OPD location is now obvious (with a mean tracking
capability of about 0.75), even though the initial signal was very
faint.}
\label{fig:typicalresultpic}
\end{figure}

We next compute the square modulus of each element of
$F_2$, i.e.\ we compute $|F_2(l_p)|^2$
and look for the peak value of the
modulus. The value of $l_p$ corresponding to the peak modulus is our
estimate of the \opd, hereafter the ``group delay estimate''.
Typical examples of the form of $|F_2(l_p)|^2$ are shown in figure~\ref{fig:typicalresultpic}.

More
sophisticated means of determining the group delay, for example
computing a centroid of the peak or making use of the phase of
$F_2(l_p)$, are not considered further here. In general such techniques
improve the accuracy of the group-delay estimate in 
high signal-to-noise-ratio (SNR) conditions, but make little
difference in the low-SNR conditions which are of most interest here.

\subsection{Coherent and incoherent integration}
\begin{figure}
\includegraphics[width=\linewidth]{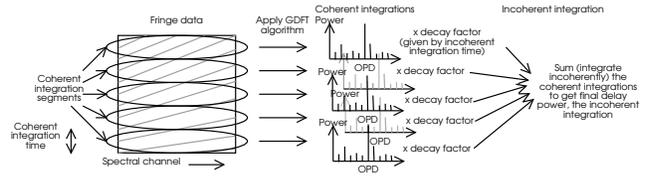}
\caption{A schematic diagram showing how a path modulation sweep is
broken into smaller lengths (coherent integrations) and the fringe
tracking algorithm is applied to each of these.  The resulting power
is then multiplied with a decay factor and summed with the power
spectrum from previous coherent integrations to give the \opd signal.
The decay factor is determined by the incoherent integration time
using a low-pass filter}
\label{fig:integ}
\end{figure}

Under conditions 
where the time for a single
path-modulation sweep is many coherence times
in length, it is likely to be advantageous to break the sweep into a number
of (possibly overlapping) 
intervals of order $t_0$ in length and to analyse these
separately. This can be accomplished by making the window function
$W_1(t_k)$ 
narrower than the sweep, and repeating the above analysis with the
window function starting at different points along the sweep. The
duration over which this window function is non-zero is 
hereafter called the ``coherent
integration time'' and the values of $F_2(l_p)$ resulting from a single
position of the window are termed a ``coherent integration''.  This
integration scheme is demonstrated in Fig.~\ref{fig:integ}

Under low SNR conditions we might want to make use of the information
from multiple coherent
integrations in order to improve the group delay estimate. Here, the
primary method we use is to average together the values of $|F_2(l_p)|^2$ from
multiple coherent integrations --- note that this can include
coherent integrations from multiple sweeps as well as from the same sweep. This
averaging of the modulus squared we call ``incoherent integration''.

The particular form of incoherent integration employed here uses an 
autoregressive filter which implements the discrete-time equivalent of a
first-order low-pass filter. We assume 
that the coherent integration is evaluated repeatedly with the
position of the coherent integration window incremented by $\Delta t$
each time, 
and introduce the notation that
$F_2(l_p,n)$ represents the result of the $n$th coherent integration.
The output $F_3(l_p,n)$ of the autoregressive filter is given by
\begin{equation}
F_3(l_p,n) = a |F_2(l_p,n)|^2 + (1-a) F_3(l_p, n-1) 
\label{sweepadd}
\end{equation}
where the decay constant $0 < a \leq 1$ is defined by
$a=1-\exp(-\Delta t/\tau)$ where
$\tau$ is the time constant of the filter, i.e.\ the effective
incoherent integration time. The position of the peak in $F_3(l_p, n)$ then
gives us our group delay estimate at time $n\Delta t$.  

It is possible  to use {\em a~priori} knowledge of the \opd offset (for
example knowledge of the fact that 
offsets at closely-spaced intervals of time are likely to be
correlated) to improve the estimate of the \opd \citep{padilla}, 
but we do not make use of such algorithms here.

The estimate of the \opd error can be used to adjust the internal
delays in the interferometer to so as to  remove any \opd
errors a closed-loop servo system. Further details
are given in \citet{npoi}.  In this paper we do not consider issues
relating to the performance of such a servo loop as a whole, 
rather we concentrate on techniques
for maximising the ability to determine the \opd, i.e.\ maximising the
fidelity of the \opd error signal.

\subsection{EMCCD output processing}
As discussed in section \ref{l3ccdsect}, an EMCCD can be used as a
GDFT detector.  However, stochastic noise is introduced into the EMCCD
output signal during the on-chip gain process in addition to the
photon shot noise expected from an ideal detector.
We process the EMCCD output using the thresholding techniques
described by \citet{basden} and \citet{basden2} to estimate the
instantaneous 
intensity $I_{jk}$. We also investigate the effect of removing this
thresholding and using instead the ``raw'' output of the
EMCCD as our estimate of $I_{jk}$.

\section{Monte-Carlo simulations}

We investigated the effects of interferometer
imperfections and processing techniques on the tracking capability of
an interferometer using a series of Monte-Carlo simulations as described below.

\subsection{Interferometer model selection}
The first step in the simulation was to select an interferometer from
the space of possible interferometer models, based on a
``prototype'' interferometer model. The prototype model had the
following characteristics:
\begin{description}
\item[Wavelength range:] 650-1000~nm
\item[Number of spectral channels:] 200
\item[Spectral sampling:] Uniform in $\sigma$ (i.e.\ a grism)
\item[Path-length modulation waveform:] Ideal (sawtooth)
\item[Peak-to-peak path-length modulation:] 60~$\mu$m
\item[Samples per sweep:] 500
\item[Atmospheric coherence time ($t_0$):] 50 sample times at
  $\lambda=825$nm
\item[Mean light level:] 0.01 photons per sample per 
pixel (100 photons detected
per $t_0$)
\item[Fringe visibility:] 0.2
\item[Intrinsic fringe phase:] 0.1
\item[Detector:] Idealised photon-counting array
\item[Static \opd error $l_\mathrm{static}$:] 25~$\mu$m
\item[Differential atmospheric path:] 0~m
\item[Water vapour partial pressure:] 0~mbar
\item[Temperature and pressure :] $pT_s/p_sT=1$
\end{description}
In the interferometer models selected, at most
one parameter was changed with respect to the prototype i.e.\ the models
selected were offset from the prototype along one ``axis'' in
the interferometer parameter space. 
In this way, the effect of
the varying the individual model parameters was kept separate.

\subsection{Fake fringe generation}
For a given interferometer model, a set of simulated spectroscopic
interferometric fringes were generated.  A Kolmogorov-Tatarski
atmospheric model phase screen was assumed with one atmospheric phase
sample for each fringe sample, and the telescope aperture size was
assumed infinitesimal so that there was no phase variation across the
aperture.  The bandpass of each spectral channel was assumed to take
the form of a top-hat function, and the finite coherence length of
each spectral channel was taken in to consideration by using a sinc
function to envelope the fringe pattern in each spectral channel.  The
mean (theoretical) intensity of the fringe pattern was computed for
each sample, and this intensity then ``Poissonised'' to produce an
integer number of photons, corresponding to the theoretical intensity,
with the introduction of photon noise.  The mean number of Poissonian
photons in each spectral channel was assumed identical to other
channels unless stated otherwise.  Telescope tip-tilt correction was
assumed perfect.

\subsection{Fringe processing}
The simulated fringes were processed to obtain an estimate for the \opd using
the algorithms described in section \ref{sec:opdestimation}.
The values of the parameters of the 
estimation algorithm were based on a prototype
algorithm with the following characteristics:
\begin{description} 
\item[Window functions $W_1$, $W_2$:] Top-hat
\item[Coherent integration time:] 100 samples
\item[Offsets between successive coherent integrations $\Delta t$:]
100 samples
\item[Incoherent integration time constant $\tau$:] 2000 samples
\item[Scale factor $s$:] 0.35
\item[Number of trial delays $2N_p$:] 200
\end{description}
Parameters of the data reduction algorithm were varied one at a time
in the same way as the interferometer models were selected.

\subsection{Performance scoring}
The difference between the estimated and ``true'' static \opd error
$l_\mathrm{static}$ was used to determine whether fringe location was
successful.  We quantify our results in terms of a ``tracking
capability'' $p_\mathrm{track}$, defined as the probability that the
estimated \opd is within $1/\Delta\sigma$ of the true \opd error. If
$p_\mathrm{track}$ is not unity, this means that for some fraction of
the time a spurious noise peak is misidentified as being the group
delay peak (since the width of the delay peak is approximately
$1/\Delta\sigma$). If the position of this spurious noise peak is used
to send corrections to a path compensation system, then the system
will tend to wander randomly away from the zero-\opd position rather
than converge towards it. Thus if $p_\mathrm{track}$ falls below
some threshold it is likely that the fringe signal will be lost.

The tracking capability was determined by repeating the simulated fringe
generation a large number of times (typically 1000-10000) 
with different random realisations of the atmospheric and photon
noise processes, processing the simulated fringe data and 
determining the relative frequency of peak
misidentification.

\section{Results}

\subsection{Importance of correct coordinate transformation}
Previous GDFT algorithms \citep{lawson} assume that the dispersion is
linear in $\sigma$ and that the sweep modulation function is
sawtoothed in shape.  These algorithms then result in poor fringe
tracking performance when applied to non-ideal situations typically
found with real interferometers as demonstrated in
tables~\ref{table:gradienttable1} (sweep shape)
and~\ref{table:gradienttable2} (dispersion type).  The low tracking
capability shown here for non-ideal situations using the previous
methods means that a much higher light level is required to achieve
the same tracking capability that is obtained using our method
presented here.

\begin{table}
\caption{A table showing the effect of sweep shape on tracking
capability when using fringe tracking algorithms from ``previous''
authors (which assume an ideal fringe pattern) and when using the
algorithm presented here (``us'').  A mean light level of 0.02 photons
per pixel and a top-hat windowing function have been used here, and
other windowing functions can lead to an improvement in tracking
capability (see table~\ref{table:windowtable2}).  The poor performance
of the previous tracking algorithms which assume a perfect fringe
pattern means that a much higher light level would be required to
achieve the same tracking capability.}
\label{table:gradienttable1}
\begin{tabular}{r|l}\hline
\opd modulation & Dispersion type\\
shape & Linear\\ \hline
Linear (Sawtooth)   & $0.88\pm0.01$\\
Sinusoidal (previous)& $0.03\pm0.01$\\
Sinusoidal (us)&$0.50\pm0.03$\\ \hline
\end{tabular}\\
\end{table}

\begin{table}
\caption{A table showing the effect of spectral dispersion on tracking
capability when using fringe tracking algorithms from ``previous''
authors (which assume an ideal fringe pattern) and when using the
algorithm presented here (``us'').  The tracking capability is not
particularly sensitive to the dispersion (for realistic dispersions on
an interferometer) when our generalised algorithm is used.  A mean
light level of 0.02 photons per pixel has been assumed here.}
\label{table:gradienttable2}
\begin{tabular}{r|l}\hline
Dispersion type & Tracking capability\\
and algorithm &\\ \hline
Linear (us and previous)& $0.88\pm0.01$\\
Prism (previous)&$0.33\pm0.03$\\
Prism (us)&$0.87\pm0.02$\\
Grating (previous)&$0.08\pm0.02$\\
Grating (us)&$0.90\pm0.01$\\ \hline
\end{tabular}\\
\end{table}

The tracking capability when using the algorithm presented here is
seen to be little affected by the spectral dispersion, with
table~\ref{table:gradienttable2} comparing the tracking
capability achieved with dispersion linear (in $\sigma$), dispersion
expected from a SF18 equilateral prism (such as used at
the COAST), and a diffraction grating with 800 lines per mm.  Other
non-realistic dispersions (e.g.\ sinusoidal) are likely to have a
greater affect on tracking capability, though are not relevant for
common dispersing elements such as those available at the COAST.

The sweep shape is seen to have a larger effect on tracking
capability, and even when using our generalised algorithms, the
tracking capability can be reduced by as much as a third when a
top-hat windowing function is used.

\subsubsection{Importance of the gradient weighting function}
We have investigated the effect on tracking capability of the gradient
function $\left.\frac{\rm d t }{\rm d
l_\mathrm{mod}(t)}\right|_{t_k}$, by including (the mathematically
correct approach) and not including (a partial coordinate transform)
this function in our \opd estimation algorithm.

\begin{figure}
\includegraphics[width=8cm]{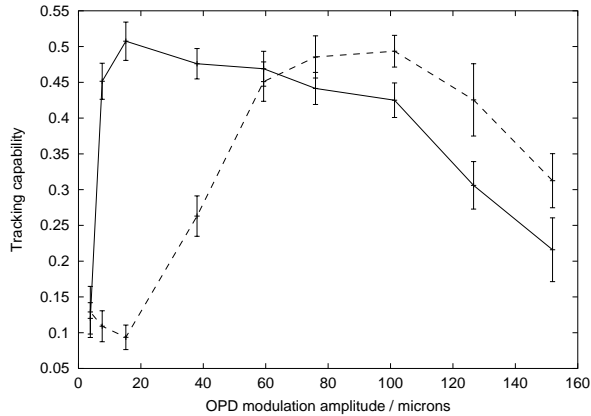}
\caption{A plot showing how a reduction of the sweep
amplitude while keeping the number of sampling points constant (all
sweeps take the same amount of time regardless of length) results in
more sample points across a fringe, and the gradient weighting
correction ($\left.\frac{\rm d t }{\rm d
l_\mathrm{mod}(t)}\right|_{t_k}$) becomes more important.  Here, the
black trace is for a fringe analysed using a gradient weighting
correction, while the dashed fringe is without, with the sweep
modulation function being sinusoidal in both cases.  When the total
\opd modulation distance is small, the weighting correction becomes
important.  However, at larger \opd modulation amplitudes (60~$\mu$m
is typical at the COAST), using the gradient weighting makes little
difference to tracking capability.  At even greater amplitudes, fringe
data is under-sampled (less than two samples per fringe), and so
tracking capability falls.}
\label{fig:stepsize}
\end{figure}

When the total \opd modulation amplitude is large (with a larger
change in the \opd between each sample, meaning that each fringe is
sampled at only a small number of points) the lack of the gradient
function does not reduce the tracking capability as shown in
Fig.~\ref{fig:stepsize}.  However as the \opd modulation amplitude is
reduced, the tracking capability begins to fall when the gradient
function is not used, since there are then more samples taken across
each fringe and so the fringe tracking algorithm is more sensitive to
the change in fringe frequency as adjacent fringes are then no longer
in phase.  Using a gradient function then helps to reduce spectral
leakage, though only has effect when the sweep or dispersion are
non-linear.

\begin{figure}
\includegraphics[width=\linewidth]{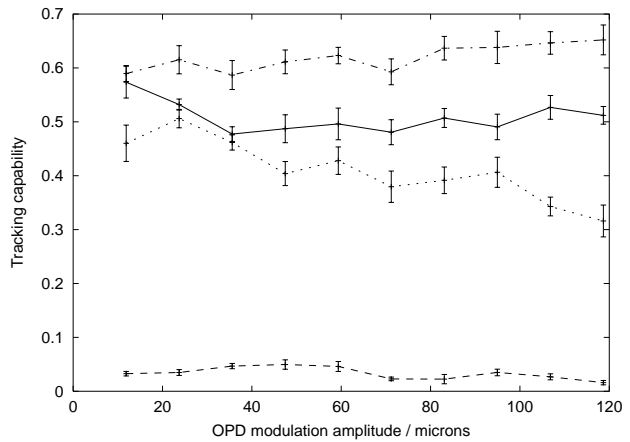}
\caption{A graph showing how tracking capability is affected by the
\opd modulation sweep size, with the number of sampling points
proportional to the sweep size (so shorter sweeps take less time), but
with the \opd changed by a constant amount between each sample,
independent of sweep size.
The dash-dotted curve shows the tracking capability for a linear
sweep.  The dashed curve shows the tracking capability that would be
obtained using algorithms assumed by previous authors for a sinusoidal
sweep.  The solid curve shows fringe data with a sinusoidal sweep
analysed with the algorithm presented here.  The dotted curve is
similar to the solid curve but does not include the gradient function
in the fringe tracking algorithm.  The use of a gradient function is
therefore advised to avoid a reduction in tracking capability.}
\label{fig:sweepsize}
\end{figure}

We also find that the tracking capability depends on the sweep size
(total distance moved during the \opd modulation cycle) with a
constant distance moved between samples (i.e.\ taking more or less
samples per sweep) when the gradient function is not used, as shown in
Fig.~\ref{fig:sweepsize}.  We therefore recommend that the gradient
function is always included in the calculations, as this ensures that
the coordinate transformation is correct, and can improve that
tracking capability.

\subsection{Effects of window functions}
In general, the use of non-Top-hat windowing functions can improve
tracking capability.  When coherent integration sections overlap
(e.g.\ the increment $\Delta t$ is half the coherent integration time
and so each coherent integration contains fringe signal from the
preceding and following coherent integrations), windowing in the time
domain can become advantageous as shown in
table~\ref{table:windowtable2}.  Even though the windowing function
reduces the importance of flux at the edge of the window, and hence
reduces the signal, the benefit from the reduction in spectral leakage
has a greater effect than the signal reduction since all parts of the
sweep are now sampled close to the window maximum due to the
overlapping windows used.

When coherent integrations do not overlap, i.e.\ the increment,
$\Delta t$, of window position between each sweep segment is equal to
or greater than the window size, we find that there is no advantage in
using a non-top-hat windowing function in either stage of the GDFT
algorithm when the sweep is linear, as shown in
table~\ref{table:windowtable}.  This is because the benefit of
reducing spectral leakage by windowing is offset by the reduction in
flux due to the signal reduction at the window edges.  Additionally,
the noise introduced by aliasing is small compared with other sources
of noise (photon shot, atmospheric) when the light level is low.

We find that the use of a top-hat windowing in non-ideal situations
(particularly with a non-linear sweep) can reduce the tracking
capability by as much as a third as shown in
tables~\ref{table:gradienttable1} and~\ref{table:gradienttable2}.
Fortunately as shown in table~\ref{table:windowtable2}, appropriate
window functions can help restore the tracking capability to that of
an ideal sawtoothed modulation and so when the tracking capability is
less than unity, it is advisable to use a non-top-hat windowing
function.  There is some evidence that a Welch window will give the
best improvement in performance.

\begin{table}
\caption{A table presenting the effect of temporal windowing functions
on tracking capability (values in the table) with non-overlapping and
overlapping coherent integration periods, comparing linear and
non-linear sweeps.  It is clear that non-Top-hat windowing functions
can help improve the tracking capability particularly if the sweep is
not linear (sawtoothed).  Uncertainties in these values are about
0.03.}
\label{table:windowtable2}
\begin{tabular}{r|llll}\hline
Sweep&\multicolumn{2}{c}{Overlapped}&\multicolumn{2}{c}{Not overlapped}\\
Window   & Linear & Sinusoidal&Linear&Sinusoidal\\ \hline
Top hat  & 0.70 & 0.59 & 0.64 & 0.44\\
Welch    & 0.71 & 0.73 & 0.59 & 0.58\\
Bartlett & 0.68 & 0.71 & 0.56 & 0.51\\
Hann     & 0.69 & 0.68 & 0.46 & 0.48\\
Hamming  & 0.72 & 0.69 & 0.53 & 0.56\\
Blackman & 0.67 & 0.65 & 0.47 & 0.43\\ \hline
\end{tabular}\\
\end{table}

\begin{table}
\caption{A table demonstrating the effect of windowing functions on
tracking capability (the values in the table) when using the \opd
estimation algorithm described here with no overlap between coherent
integration periods.  These values were obtained using a mean light
level of 100 photons per $t_0$, for a linear sweep and dispersion.
Errors in these tracking capability values are about 0.03 and so it is
clear that the use of a top-hat windowing function gives best tracking
capability.}
\label{table:windowtable}
\begin{tabular}{r|l|l}\hline
Window type & Sweep window & Spectral window\\ \hline
Top hat & 0.63 & 0.63\\
Welch & 0.58 & 0.58\\
Bartlett & 0.51 & 0.53\\
Hann & 0.52 & 0.50\\
Hamming & 0.56 & 0.49\\
Blackman & 0.46 & 0.40\\ \hline
\end{tabular}
\end{table}

\subsection{Longitudinal dispersion}
\label{sect:longdisp}
\begin{figure}
\includegraphics[width=8cm]{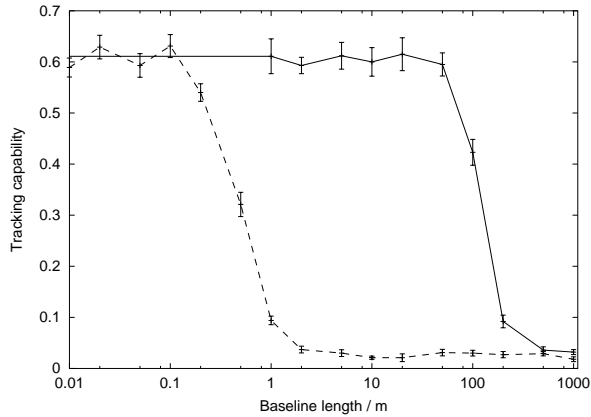}
\caption{The effects of longitudinal dispersion and a non-zero
atmospheric path difference on tracking capability when path
compensation is not carried out in a vacuum.  The dashed curve shows
the effect of longitudinal dispersion when using previous fringe
tracking algorithms \citep{lawson} while the solid curve shows
the effect when using the algorithm presented here.  The improvement
for longer baselines is clear.}
\label{fig:longdispersionpic}
\end{figure}

If the distance travelled from a star along two separate light paths
through the atmosphere before recombination differs by a large amount,
then the phase difference introduced by atmospheric longitudinal
dispersion (Eq.~\ref{eq:longitudinal}) between two well spaced
spectral channels will be much greater than $2\pi$ (the atmospheric
dispersion is seen to introduce a phase difference of about 13 radians
or 2~$\mu$m per metre of differential air path between 650 and
1000~nm).  If this phase difference is not corrected for in the fringe
tracking algorithms (this has been the case for most previous
authors), the tracking capability will fall quickly with projected
baseline length even for short baselines as shown in
Fig.~\ref{fig:longdispersionpic}.

When a correction is made for this longitudinal dispersion phase
(using Eq.~\ref{eq:longitudinal}), the tracking capability
will be reduced as the phase difference approaches $2\pi
l_\mathrm{coh}/\lambda$ (the phase of the coherence length of the
light in any given pixel), since it becomes impossible for fringes to
be detected simultaneously in two well spaced wavelengths.
Fig.~\ref{fig:longdispersionpic} shows this effect, with
tracking capability becoming reduced for atmospheric path differences
greater than about 100~m, though as the light level or visibility are
increased, longer baselines can still be used since the fringe signal
is then clearer even though separated by large longitudinal dispersion
phases.  Implementing vacuum delay lines in the interferometer removes
any differential longitudinal dispersion, as the total distance
travelled through the atmosphere is then identical for each light beam
when the \opd is zero.

\subsubsection{Atmospheric uncertainties}
\begin{figure}
\includegraphics[width=8cm]{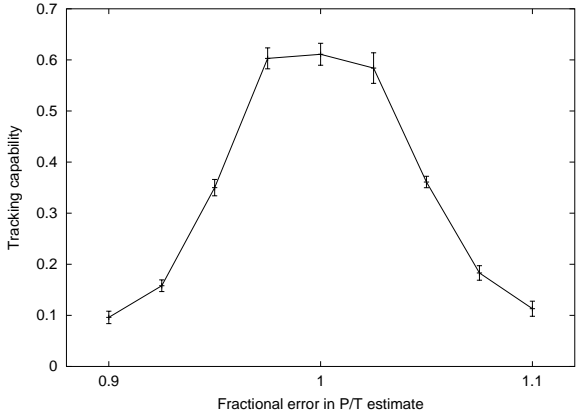}
\caption{Simulated effect of imprecise knowledge of atmospheric
temperature and pressure when fringe tracking for a projected baseline
length of 10m, showing that the reduction in tracking capability is
approximately Gaussian about the true value, with the width
corresponding to an error of about 0.07 in pressure or temperature.
This effect is due to the atmospheric longitudinal dispersion.}
\label{impreciseatmos}
\end{figure}

The exact form of the longitudinal dispersion may not be known, being
dependent on both the model used and the atmospheric conditions.  We
therefore investigate the effect of imprecise knowledge of atmospheric
conditions when interpreting the results, and also investigate the
effects of model dependencies.  Fig.~\ref{impreciseatmos}(a) shows
the effect on tracking capability when different atmospheric
conditions are used to produce fringes ($(p/T)_\mathrm{produce}$) than
are assumed when computing the \opd ($(p/T)_\mathrm{compute}$), and
this uncertainty can be characterised by the quantity
\begin{equation}
C=\frac{(p/T)_\mathrm{produce}}{(p/T)_\mathrm{compute}}.
\end{equation}
The loss in tracking capability is only small if $|1-C|<0.02$.
This loss is found to depend on $|1-C|$ following a Gaussian distribution
with a width of about 0.07.  For example, if a fringe pattern is
simulated at standard temperature and pressure, and the fringe
tracking algorithm computed assuming a pressure increase of one
percent ($|1-C|=0.01$), the tracking capability is barely affected.
However, if the pressure is increased by ten percent ($|1-C|=0.1$), the
tracking capability is greatly reduced.

\begin{figure}
\includegraphics[width=8cm]{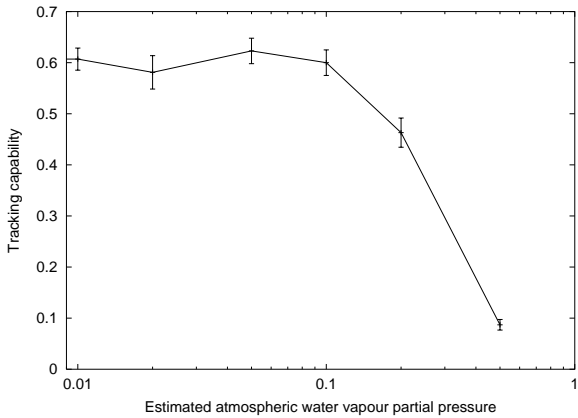}
\caption{Simulated effect of imprecise knowledge of atmospheric water
partial pressure when fringe tracking for a differential atmospheric
path of 10m, showing that the reduction in tracking capability is
minimal for all likely atmospheric humidities.  No water vapour was
assumed when producing the fringes.}
\label{imprecisepW}
\end{figure}

We also find that the tracking capability is virtually independent of
the error made when estimating humidity for atmospheric conditions
found on earth (Fig.~\ref{imprecisepW}).


We have yet to discover how effective our method will be when applied
to real data, though it is likely that any model discrepancy will be
small, since the atmospheric refractive index has been well studied.
Atmospheric conditions are typically known to much better than one
percent, and so it is likely that this effect will be minimal.  The
longitudinal phase correction does not depend on terms independent of
wavelength, since this just adds a constant phase.

\subsection{Pixel-wavelength relationship}
\begin{figure}
\includegraphics[width=8cm]{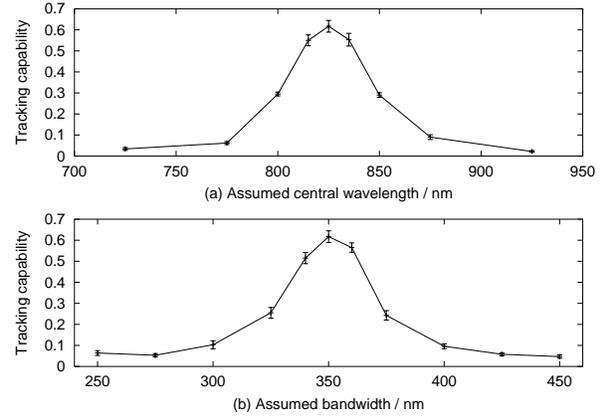}
\caption{Plots showing the effect of an unknown wavelength-pixel
relationship.  In (a), we can see that it is important to be able to
locate the central wavelength of the dispersed light accurately to
within 10-20 nm to avoid a significant loss in tracking capability.
In (b), we see that the bandwidth of the light must be estimated to
within about 20 nm.  In both of these plots, the true total bandwidth
was 350 nm, centred at 825 nm.  We find that this tolerance is
independent of the dispersion (e.g.\ linear, prism or grating).}
\label{pxlwavelengthreln}
\end{figure}

We find that if the pixel-wavelength relation is not known precisely,
then the tracking capability is reduced.  Fig.~\ref{pxlwavelengthreln}
(a) shows that the centre of the spectrum should be located with about
1-2 percent accuracy (typically 5-10 pixels for a 350nm bandwidth
spectrum at 820nm, assuming 256 spectral channels).  Likewise,
Fig.~\ref{pxlwavelengthreln} (b) shows that the total bandwidth should
be known to about 5 percent so that tracking capability is not
significantly reduced.  In practice, the wavelength-pixel relationship
will be known to better than this as it can be accurately determined
from the fringe frequencies, and so the effect shown here will
be minimal unless misalignment occurs.

\subsection{Sampling scale}

\begin{figure}
\includegraphics[width=8cm]{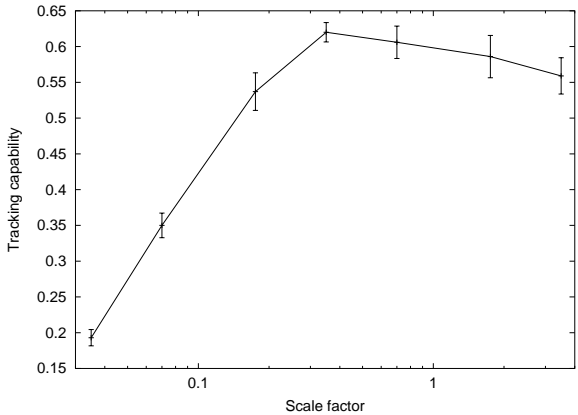}
\caption{Showing the effect of changing the scale factor.  If the
scale factor is too small, the tracking capability is reduced, while
if too large, the resolution to which the \opd can be specified is
reduced.  We see that tracking capability remains high for scale
factors between 0.2-2.  However, if the scale factor gets too large,
we are unable to identify the \opd correctly, as our resolution is then
not great enough.}
\label{scalepic}
\end{figure}

The sampling scale ($s$) used does not affect the way the detector is
read out and hence many different scales can be used each modulation
half-period until the scale giving maximum tracking capability is
found.  As shown in Fig.~\ref{scalepic}, we find that when the
sampling scale is large ($>2$) or small ($<0.2$), tracking
capability may be reduced since there is an increase in spectral
leakage.  By using a small sampling size, it is possible to locate a
fringe envelope maximum to within a fraction of a wavelength provided
the \opd is small.

If computing power limits the number of delay space samples that can
be calculated for each data frame, it may be advantageous to use a
large scale factor to identify the \opd initially.  The \opd can then be
reduced using optical path compensators, and a smaller sampling scale
can then be used to refine the \opd estimation, whilst all the time
calculating a constant number of delay space samples.

\subsection{Coherent and incoherent integrations}

It has been shown \citep{buscher} that the optimal time over which to
integrate coherently is about $1.6{\rm t_0}$, allowing for a tradeoff
between received energy and temporal smearing due to atmospheric
perturbations.  Further incoherent integration
(Eq.~\ref{sweepadd}) is of advantage, where
frames are summed over a time-scale of typically between $20\rm{t_0}$
\citep{lawson} - $40{\rm t_0}$ \citep{buscher}.  There is some
discrepancy between these two estimates, and so we investigate the
best time-scale over which to integrate both coherently and
incoherently using the sweep combination algorithm, Eq.~\ref{sweepadd}.

\begin{figure}
\includegraphics[width=8.5cm]{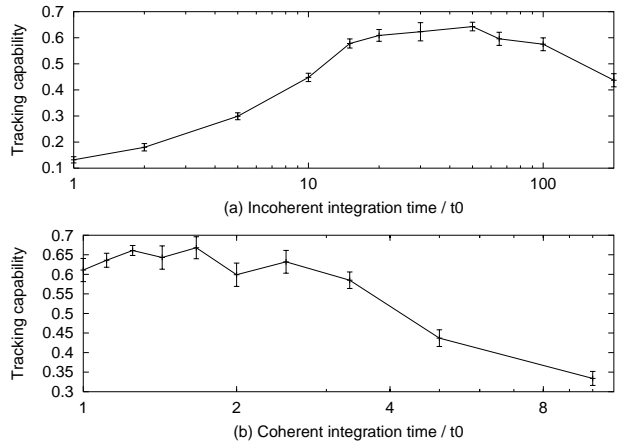}
\caption{The effect of integration times on tracking capability.  The
top trace shows the effect of incoherent integration time on tracking
capability, and is seen to peak between about $20-50t_0$.  The lower
trace shows the tracking capability as a function of coherent
integration time, and shows that shorter integration times are favoured.}
\label{integtime}
\end{figure}

Fig.~\ref{integtime} shows the simulated effect of integration time on
tracking capability.  Short coherent integration time-scales of about
$t_0 - 2t_0$ achieve the best tracking capability during a single
sweep, in agreement with \citet{buscher}, while using
an incoherent integration time of between 20-50~$t_0$ also maximises
the tracking capability.

We recommend the general use of a coherent integration time of
1.6~$t_0$ and an incoherent integration time of about 30~$t_0$
for general use.

\subsection{Effects of intensity variation with wavelength}

Intrinsic variations of source intensity with wavelength (for example
due to stellar atmosphere absorption), and a change in detector
quantum efficiency with wavelength mean that the mean intensity of a
fringe will depend on spectral channel.  For most of our calculations
we assume a flat source power spectrum, with equal power produced for
each spectral channel.  However, we have also investigated several
other models for stellar atmospheres, particularly those for cool
stars, as these are commonly observed with optical stellar
interferometers.  We use a stellar power spectrum for both 61 Cyg A
and WX Cam taken from \citet{turnshek}.  We also investigate the
inclusion of a model of EMCCD quantum efficiency and find that using
such spectra makes negligible difference to the fringe tracking
capability, when compared with models using a constant mean flux with
the same number of photons detected in the fringe.  This is the case
even when large absorption bands are present within the stellar
spectra.

We find that for a given mean detected flux, the tracking capability
does not depend on the wavelength intensity relationship for commonly
occurring wavelength-intensity dependencies.  The presence of large
stellar atmosphere absorption bands have little effect on tracking
capability.

\subsection{Combination of spectral channels}
At low light levels, there may only be at most a few detected photons
in each spectral channel during the optical delay modulation
half-period.  It may therefore be tempting to combine spectral
channels together, with the hope of reducing the phase error when
estimating the phase of the fringe in each spectral channel.

Spectral channel combination can either be carried out before CCD
readout, or afterwards in software.  Pre-readout spectral channel
combination has the disadvantage that once spectral channels have been
combined there is no way of returning to the higher spectral
resolution.  Therefore, spectral combination in software is the best
option if CCD readout noise is negligible, as with EMCCDs.  If CCD
readout noise is non-negligible or a very rapid readout is required,
it may be appropriate to bin spectral channels together before
readout.

\begin{figure}
\includegraphics[width=8cm]{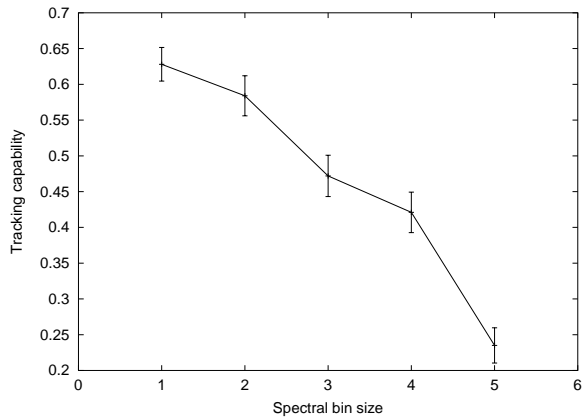}
\caption{A graph showing how combining spectral channels by summing
raw data in adjacent channels can reduce tracking capability due to
smearing of fringe, and so is not recommended.}
\label{binnedspecchan}
\end{figure}

Combining spectral channels will reduce the coherence length of each
channel, and so if the number of spectral channels combined becomes
large, there will be a reduction in tracking capability since as the
coherence length approaches the \opd, it becomes more difficult to
find and track fringes.  We investigate the effects of combining
between 1-5 spectral channels together, and find that there is a
decrease in tracking capability as more channels are combined
(Fig.~\ref{binnedspecchan}), due to smearing of the fringes at the
edges of the modulation where the rate of change of fringe phase with
wavenumber typically becomes large.  This technique is therefore not
recommended.

\subsection{EMCCDs for group delay fringe tracking}
Being array detectors, EMCCDs can be used as group delay fringe
tracking detectors, since they are capable of recording a one
dimensional spectrum rapidly.  We model the use of an EMCCD for GDFT
assuming the model given by \citet{basden}, with a mean on-chip gain
of 6629, assuming a RMS readout noise of 50 electrons is introduced to
the multiplied signal.  The EMCCD output first has a noise level at
300 electrons removed, and is then treated using different
thresholding strategies following \citet{basden,basden2} (the
Analogue, Digital, Uniform and Poisson Probability thresholding
strategies), ignoring the effects of dark current which will be small
due to the fast readout.

\begin{figure}
\includegraphics[width=8cm]{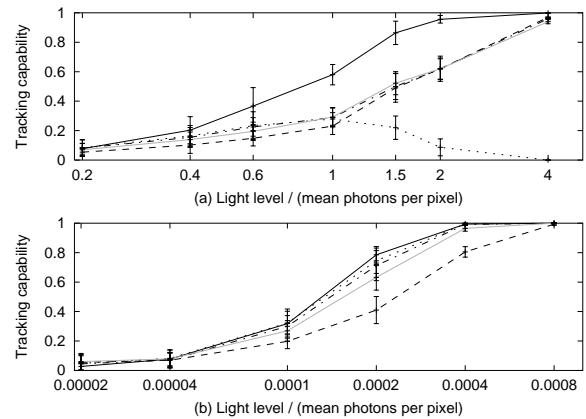}
\caption{Simulated tracking capability when using an EMCCD as the
detector for a fringe visibility of (a) 0.02 and (b) unity.  The black
curve represents the tracking capability for a perfect detector
(Poisson noise only).  The dashed curve is when using the raw output
of an EMCCD, the dotted curve is when using a single threshold on the
raw EMCCD output, the dot-dashed curve is when using a uniform
thresholding strategy and the grey curve is when using a Poisson
probability thresholding strategy.  At low light levels, there is much
advantage of thresholding the data.  When the light level is high,
typically twice as much light is required to achieve the same tracking
capability with the raw output as would be needed for a perfect
detector.}
\label{fig:l3pic}
\end{figure}

Fig.~\ref{fig:l3pic} shows the effect that the EMCCD and associated
thresholding strategies have on the tracking capability of a GDFT
system.  If the Analogue strategy is used, twice as many photons are
required to achieve the same tracking capability as that achieved with
a perfect detector (independent of the light level), and this is
expected due to the reduction in effective quantum efficiency due to
the stochastic multiplication process \citep{basden}.  However, if the
EMCCD output is processed appropriately, the effect of stochastic
multiplication noise can be minimised at low light levels, and the
fringe tracking capability can be improved as shown in these figures.
At light levels greater than about two photons per pixel
(Fig.~\ref{fig:l3pic}(a)), there is little benefit in thresholding the
EMCCD output, in agreement with results obtained in previous chapters.
The Uniform thresholding strategy \citep{basden2} appears to give the
most consistent improvement in tracking capability, and at light
levels less than about 0.2 photons per pixel the tracking capability
is almost that of an ideal detector since the EMCCD is then operated
in photon counting mode.

\subsubsection{Low light levels}
At low light levels ($<1$ photon per pixel) the EMCCD can be operated
in a photon counting mode, treating every signal above some noise
level as representing one photon \citep{basden}.  Tracking capability
is almost that of a perfect detector (Fig.~\ref{fig:l3pic}(b)).
Alternatively using the Uniform multiple thresholding strategy
described by \citet{basden2} gives a similar tracking capability.

\subsubsection{Higher light levels}
At higher light levels (with the mean light level approaching or
greater than one photon per pixel), coincidence losses when using a
single threshold become large, since there is a significant
probability that we would interpret two or more photons as being one
photon.  This leads to a reduction in tracking capability as
demonstrated in Fig.~\ref{fig:l3pic}(a).

Using multiple thresholds \citep{basden2} can improve the tracking
capability at these light levels, and although the estimated number of
photons is not always correct, we do significantly better than when
using a single threshold.  This leads to an improvement in tracking
capability, though we are still not able to track as well as an ideal
detector would allow.  The results in Fig.~\ref{fig:l3pic} show that we
are still able to improve tracking capability over that achieved with
the raw output for light levels up to about 2 photons per pixel.

Using the raw EMCCD output (after subtracting a noise level) gives a
better result than when using a single threshold for light levels
above about 1 photons per pixel.  It is also not significantly worse
than a multiple thresholded output mode when the light level is
greater than about two photons per pixel.  At these light levels,
approximately twice as much light is required to achieve a given
tracking capability as would be needed using an ideal detector, which
is the main disadvantage when using EMCCDs at high light levels.

\subsubsection{Ideal output interpretation}
These results are in agreement with those presented by \citet{basden}
showing that the effect of EMCCD stochastic multiplication noise can
be successfully reduced at low light levels.  The best approach for
fringe tracking with an EMCCD is to place the EMCCD output into
multiple thresholds with a threshold spacing dependent on mean gain
and light level (which can be estimated for each spectral channel by
averaging the received signal), as described by \citet{basden2}.  This
mode allows us to track well at any light level, making most use of
the information that we have.  At the lowest light levels, we have a
tracking capability similar to that of an ideal detector, while at
higher light levels (of order one photon per pixel or more) it is not
so good.

Since best results are achieved at light levels less than one photon
per pixel, the EMCCD should be operated in this regime if possible,
increasing the pixel rate where appropriate to reduce the signal on
each pixel.  At very high light levels, the EMCCD could be used in a
non-amplifying mode, where the signal detected by the CCD is not
amplified at all, and noise in the signal is then due to readout noise
and photon Poisson noise.  However, it is unlikely that spectroscopic
interferometric signals will contain this amount of flux.

\section{Conclusions}
The maximisation of the performance of GDFT algorithms using
Monte-Carlo simulation has been investigated.  Previous authors
(\citet{lawson} and references therein) have typically assumed
ideal conditions when investigating GDFT algorithms.  However, here
it has been shown that when used in real situations, these algorithms
may give poor performance.  Investigations into the performance of a
generalised GDFT algorithm has shown that the tracking capability can
be improved, leading to the following conclusions:
\begin{enumerate}
\item The non-linear dispersions typically resulting from prism or
grating dispersion elements do not affect tracking capability
significantly.  When appropriate windowing functions are used, a
non-sawtoothed \opd modulation will not affect the tracking
capability, though this can be reduced by up to a third if
a top-hat windowing function is used,

\item Atmospheric longitudinal dispersion can in theory be corrected
for sub-kilometre baselines, though analysis of real data is needed to
verify this.

\item Coherent integration for about $1.6 t_0$ gives the largest
improvement in tracking capability.

\item Incoherent integration times between about 20-50~$t_0$ lead to
maximisation of the tracking capability.  If longer incoherent
integration times are used, the \opd signal can become smeared over
several locations due to the temporal atmospheric fluctuations.

\item Windowing functions can lead to an improvement in tracking
capability particularly when used with overlapping coherent
integrations, and when the sweep function is non-ideal.  When
performing a transform over spectral channels, spectral windowing
functions are not seen to give a significant improvement in tracking
capability.

\item Summing adjacent spectral channels leads to a reduction in
tracking capability.

\item Altering the sampling scale of which the delay space is sampled
can be used to increase the range of \opd positions detectable, or
increase the accuracy of detection.

\item Using an EMCCD as a photon counting detector gives the tracking
capability expected from an ideal detector at light levels less than
one photon per pixel.  At higher light levels, up to twice as much
light needs to be collected on an EMCCD to achieve the same tracking
capability as an ideal detector.

\end{enumerate}

\bsp
\label{lastpage}
\end{document}